\documentclass[12pt]{article}
\usepackage{iopconf}
\usepackage{graphicx}
\usepackage{bm}

\begin{document}

\title{Helicity Basis for Spin 1/2 and 1, and Discrete Symmetry Operations}

\author{Valeri V. Dvoeglazov \dag \ and J. L. Quintanar Gonz\'alez \dag}

\affil{\dag\ Universidad de Zacatecas\\ A. P. 636, Suc. UAZ, C. P. 98062, Zacatecas, Zac., M\'exico\\ 
E-mail: valeri@cantera.reduaz.mx, el\_leo\_xyz@yahoo.com.mx}

\beginabstract
We study the theory of the $(1/2,0)\oplus (0,1/2)$ and $(1,0)\oplus (0,1)$
representations of the Lorentz group in  the helicity basis. The helicity eigenstates
are {\it not} the parity eigenstates. This is in accordance with the
idea of Berestetski\u{\i}, Lifshitz and Pitaevski\u{\i}. The properties
of the helicity eigenstates with respect to the charge conjugation and the $CP$- conjugation are also considered.
\endabstract

\section{Introduction.}

What are motivations for this work? First of all, Berestetski\u{\i}, Lifshitz and Pitaevski\u{\i} stressed~\cite{Lan}: ``... the orbital angular momentum ${\bf
l}$ and the spin ${\bf s}$ of a moving particle are not separately
conserved. Only the total angular momentum ${\bf j}= {\bf l}+{\bf s}$ is
conserved. The component of the spin in any fixed direction (taken as
$z$-axis) is therefore also not conserved, and cannot be used to enumerate
the polarization (spin) states of moving particle." Moreover, they made conclusion that the helicity eigenstates are {\it not} the parity eigenstates for any spin~\cite[p.59]{Lan}, see also~\cite{Novozh}. Next, working with 
the $\Psi_{(6)} = \mbox{column} ({\bf E}+i{\bf B},\,\, {\bf E}-i{\bf B})$
in the Weinberg-Tucker-Hammer formalism~\cite{weinbergp,tucker,valeri7}
I found that upon rotation of $\Psi$ we can obtain much more equations 
for the antisymmetric tensor (AST) field of the 2nd rank than in the accustomed
Proca formalism. Some of them imply parity-violating transitions (i.~e., contain the dual tensor and the axial-vector 4-potential). Then, we generalized the Dirac formalism~\cite{Barut,DasG,Dv2,Dv2a} and
the Bargmann-Wigner formalism~\cite{Dv3,Dv4,Dv5}. 

In this paper we are going to study transformations from the standard basis
to the helicity basis in the Dirac theory and in the $(1,0)\oplus (0,1)$ Sankaranarayanan-Good theory~\cite{SG,Dva}. The spin basis rotation
{\it changes} the properties of the corresponding states with respect to
parity. The parity is a physical quantum number; so, we try to extract
corresponding physical contents from considerations of the various spin
bases.

\section{The $(1/2,0)\oplus (0,1/2)$ case.}

We know that in the $(1/2,0)\oplus (0,1/2)$ representation the helicity operator ${\bf\sigma}\cdot
\widehat {\bf p}/2 \otimes I_2$, $\widehat {\bf p} = {\bf p}/\vert {\bf
p}\vert$, commutes with the Hamiltonian (more precisely, the commutator is
equal to zero when acting on the one-particle plane-wave solutions).
Previously, the 4-spinors have been studied very well when the spin basis has been chosen in such a way that they were eigenstates
of the $\hat {\bf S}_3$ operator, e.~g., ref.~\cite{ryder}. The
helicity basis case has not been studied almost at all
(see, however, refs.~\cite{Novozh,Grei,JW}).  The 2-eigenspinors of the helicity operator 
${1\over 2} {\bf \sigma}\cdot\widehat
{\bf p} = {1\over 2} \pmatrix{\cos\theta & \sin\theta e^{-i\phi}\cr
\sin\theta e^{+i\phi} & - \cos\theta\cr}$
can be defined as follows~\cite{Var,Dv1}:
\begin{eqnarray}
\phi_{{1\over 2}\uparrow}=\pmatrix{\cos{\theta \over 2} e^{-i\phi/2}\cr
\sin{\theta \over 2} e^{+i\phi/2}\cr}\,,\quad
\phi_{{1\over 2}\downarrow}=\pmatrix{\sin{\theta \over 2} e^{-i\phi/2}\cr
-\cos{\theta \over 2} e^{+i\phi/2}\cr}\,,\quad\label{ds}
\end{eqnarray}
for $\pm 1/2$ eigenvalues, respectively.

We\, start from the Klein-Gordon equation, generalized for
describing the spin-1/2  particles (i.~e., two additional degrees
of freedom); $c=\hbar=1$, see ref.~\cite{Gerst}.
\begin{equation}
(E+{\bf \sigma}\cdot {\bf p}) (E- {\bf \sigma}\cdot {\bf p}) \phi
= m^2 \phi\,.\label{de}
\end{equation}
It can be re-written in the form of the set of two first-order equations
for 2-spinors as in~\cite{Gerst}. Simultaneously, we observe that they may be chosen
as eigenstates of the helicity operator which present
in (\ref{de}). If the $\phi$ spinors are defined by the equation (\ref{ds}), then we can construct the corresponding $u-$ and $v-$ 4-spinors:
\begin{eqnarray}
u_\uparrow &=&
N_\uparrow^+ \pmatrix{\phi_\uparrow\cr {E-p\over m}\phi_\uparrow\cr} =
{1\over \sqrt{2}}\pmatrix{\sqrt{{E+p\over m}} \phi_\uparrow\cr
\sqrt{{m\over E+p}} \phi_\uparrow\cr},
u_\downarrow = N_\downarrow^+ \pmatrix{\phi_\downarrow\cr
{E+p\over m}\phi_\downarrow\cr} = {1\over
\sqrt{2}}\pmatrix{\sqrt{{m\over E+p}} \phi_\downarrow\cr \sqrt{{E+p\over
m}} \phi_\downarrow\cr}\,,\label{s1}\\
v_\uparrow &=& N_\uparrow^- \pmatrix{\phi_\uparrow\cr
-{E-p\over m}\phi_\uparrow\cr} = {1\over \sqrt{2}}\pmatrix{\sqrt{{E+p\over
m}} \phi_\uparrow\cr
-\sqrt{{m\over E+p}} \phi_\uparrow\cr},
v_\downarrow = N_\downarrow^- \pmatrix{\phi_\downarrow\cr
-{E+p\over m}\phi_\downarrow\cr} = {1\over
\sqrt{2}}\pmatrix{\sqrt{{m\over E+p}} \phi_\downarrow\cr -\sqrt{{E+p\over
m}} \phi_\downarrow\cr}\nonumber\\
&&\label{s2}
\end{eqnarray} 
where the normalization to the unit
($\pm 1$) was used. They satisfy the Dirac equation with $\gamma$'s to be in the spinorial representation. One can prove that the matrix
$P=\gamma^0 = \pmatrix{0&1\cr 1& 0\cr}$
can also be used in the parity operator as well as
in the original Dirac basis~\cite{valerihelic}. 
Of course, it is possible to expand the 4-spinors defined in the parity basis   in  linear superpositions of the helicity basis
4-spinors and to find corresponding coefficients of the expansion:
\begin{eqnarray}
u_\sigma ({\bf p}) &=& A_{\sigma\lambda} u_\lambda ({\bf p})
+ B_{\sigma\lambda} v_\lambda ({\bf p})\,,\\
v_\sigma ({\bf p}) &=& C_{\sigma\lambda} u_\lambda ({\bf p})
+ D_{\sigma\lambda} v_\lambda ({\bf p})\,.
\end{eqnarray}
Neither $A$ nor $B$ are unitary:
\begin{eqnarray}
A= (a_{++} +a_{+-}) (\sigma_\mu a^\mu) +(-a_{-+} +a_{--})
(\sigma_\mu a^\mu) \sigma_3\,,\\
B= (-a_{++} +a_{+-}) (\sigma_\mu a^\mu) +(a_{-+} +a_{--})
(\sigma_\mu a^\mu) \sigma_3\,,
\end{eqnarray}
where
\begin{eqnarray}
a^0 &=& -i\cos (\theta/2) \sin (\phi/2) \in \Im m\,,\quad
a^1 = \sin (\theta/2) \cos (\phi/2)\in \Re e\,,\\
a^2 &=& \sin (\theta/2) \sin (\phi/2) \in \Re e\,,\quad
a^3 = \cos (\theta/2) \cos (\phi/2)\in \Re e\,,
\end{eqnarray}
and
\begin{eqnarray}
a_{++} &=&\frac{\sqrt{(E+m)(E+p)}}{2\sqrt{2} m}\,,\quad
a_{+-} =\frac{\sqrt{(E+m)(E-p)}}{2\sqrt{2} m}\,,\\
a_{-+} &=&\frac{\sqrt{(E-m)(E+p)}}{2\sqrt{2} m}\,,\quad
a_{--} =\frac{\sqrt{(E-m)(E-p)}}{2\sqrt{2} m}\,.
\end{eqnarray}
However, $A^\dagger A+B^\dagger B =1$, so the transformation matrix ${\cal U}$
is unitary. 

We now investigate the properties of the helicity-basis 4-spinors
with respect to the discrete symmetry operations $P$ and $C$.
It is expected that $\lambda\rightarrow -\lambda$ under parity,
as in~\cite{Lan}. With respect to ${\bf p} \rightarrow -{\bf p}$ 
 the helicity 2-eigenspinors
transform as follows: $\phi_{\uparrow\downarrow} \Rightarrow
-i \phi_{\downarrow\uparrow}$, ref.~\cite{Dv1}.
Hence,
\begin{eqnarray}
Pu_\uparrow (-{\bf p}) &=& -i u_\downarrow ({\bf p})\,,
Pv_\uparrow (-{\bf p}) = +i v_\downarrow ({\bf p})\,,\\
Pu_\downarrow (-{\bf p}) &=& -i u_\uparrow ({\bf p})\,,
Pv_\downarrow (-{\bf p}) = +i v_\uparrow ({\bf p})\,.
\end{eqnarray}
Thus, on the level of classical fields, we observe that
the helicity 4-spinors transform to the 4-spinors of the opposite
helicity. The charge conjugation operation is defined as
$C =\pmatrix{0&\Theta\cr
-\Theta & 0\cr} {\cal K}$\,.
Hence, we observe
\begin{eqnarray}
Cu_\uparrow ({\bf p}) &=& - v_\downarrow ({\bf p})\,,
Cv_\uparrow ({\bf p}) = +  u_\downarrow ({\bf p})\,,\\
Cu_\downarrow ({\bf p}) &=& + v_\uparrow ({\bf p})\,,
Cv_\downarrow ({\bf p}) = - u_\uparrow ({\bf p})\,.
\end{eqnarray}
due to the properties of the Wigner operator $\Theta \phi_\uparrow^\ast =
-\phi_\downarrow$ and $\Theta \phi_\downarrow^\ast = +\phi_\uparrow$. This is similar to the textbook case.
For the $CP$ (and $PC$) operation we get:
\begin{eqnarray}
C P u_\uparrow (-{\bf p}) &=& -PC u_\uparrow (-{\bf p})
= +i v_\uparrow ({\bf p}),
C P u_\downarrow
(- {\bf p}) = - P C u_\downarrow (-{\bf p}) = -i v_\downarrow ({\bf
p}),\\
C P v_\uparrow (-{\bf p}) &=& - P C v_\uparrow (-{\bf p}) =
+  i u_\uparrow ({\bf p}),
C P v_\downarrow (-{\bf p}) = - P C v_\downarrow (-{\bf p}) =
- i u_\downarrow ({\bf p}),
\end{eqnarray}
which are different from the Dirac `common-used' case.
Similar conclusions can be drawn in the Fock space.

\section{The $(1,0)\oplus (0,1)$ case.}

In this Section we are going to investigate the behaviours of the field functions of the $(1,0)\oplus (0,1)$ representation in the helicity basis with respect to $P$, $C$ and $CP$ operations. 

Let us start from the Klein-Gordon equation written for the 3-component function ($\hbar=c=1$):
\begin{equation}
(E^2-{\bf p}^2)\psi_{(3)}=m^2\psi_{(3)}. \label{1b}
\end{equation}
The equation (\ref{1b}) can be re-written in the form:
\begin{equation}
( E-{\bf S\cdot p})(E+{\bf S\cdot p})_{ij}{\bf\psi}^j- p_i p_j{\bf\psi}^j=m^2\psi^i.
\label{2b}
\end{equation}
In the coordinate space it is of the second order in the time derivative, 
but as in the spin-1/2 case~\cite{valerihelic} we can reduce it to the set of  the 3-``spinor" equations of the first orders.  We can denote:
\begin{equation}
(E+{\bf S\cdot p}){\bf \psi}=m{\bf\xi},\, 
p^ip^j\psi^j={\bf p\,(p}\cdot{\bf\psi})=m {\bf p}\,\varphi . \label{b}
\end{equation}
Hence, the equation (\ref{2b}) is written as
\begin{equation}
m(E-{\bf S\cdot p}){\bf\xi}-m{\bf p}\,\varphi=m^2{\bf\psi}. \label{c}
\end{equation}
Now, we define ${\bf\psi}={\bf E}-i{\bf B}$. We can obtain (cf. with ref.~\cite{Dv2a}) 
\begin{eqnarray}
&& \nabla\times{\bf B}-\frac{\partial {\bf E}}{\partial t}=-m\cdot{\sf Im}({\bf\xi}), \hspace{5mm}
\nabla\times{\bf E}+\frac{\partial {\bf B}}{\partial t}=m\cdot{\sf Re}({\bf \xi}),\\
&& \nabla\cdot{\bf B}=-m\cdot{\sf Re}(\varphi)+const_x\,, \hspace{5mm}
\nabla\cdot{\bf E}=-m\cdot{\sf Im}(\varphi)+const_x\,,
\end{eqnarray}
respectively, by means of separation of the equations in (\ref{b}) into the real and imaginary parts. Next, we fix $\varphi=im\phi$ and ${\bf\xi}=im{\bf A}$, with $\phi$ and ${\bf A}$ being the electromagnetic-like potentials. The well-known Proca equation follows
$\partial_{\mu}F^{\mu \nu}+m^2 A^{\nu}=0$.
For the sake of completeness let us substitute  $\varphi$ and ${\bf \xi}$ in the equation (\ref{c}). The result is $-\frac{\partial {\bf A}}{\partial t}-\nabla \phi={\bf E}$ y $\nabla \times {\bf A}={\bf B}$, that is equivalent to the second Proca equation
$F^{\mu \nu}=\partial^{\mu}A^{\nu}-\partial^{\nu}A^{\mu}$.
We also can take the complex conjugates of the equations (\ref{b},\ref{c}) and now define  ${\bf \chi}={\bf E}+i{\bf B}$. As a result we have 
\begin{eqnarray}
(E-{\bf S\cdot p}){\bf\chi}=-m{\bf\xi}&\hspace{3mm}\mbox{or}\hspace{3mm}&
(E-{\bf S\cdot p})({\bf E}+i{\bf B})=-im^2{\bf A}, \label{aa}\\
p^ip^j\chi^j={\bf p(p}\cdot{\bf\chi})=-m{\bf p}\varphi&\hspace{3mm}\mbox{or}\hspace{3mm}& {\bf p\, [p}\cdot({\bf E}+i{\bf B})]=-im^2{\bf p}\phi, \label{bb}\\
(E+{\bf S\cdot p}){\bf\xi}-{\bf p}\varphi=-m{\bf\chi}&\hspace{3mm}\mbox{or}\hspace{3mm}&
(E+{\bf S\cdot p}){\bf A}-{\bf p}\phi=i({\bf E}+i{\bf B}), \label{cc}
\end{eqnarray}
It is possible to put the above equations in the Kemmer 
$10\times 10$ matrix form (cf.~\cite{greiner}). The equation contains the part corresponding to the 4-vector potential and to fields. 
Taking into account the Proca equations, the deinitions of ${\bf E}^i= F^{i0}$, ${\bf B}^i=-{1\over 2}\epsilon^{ijk} F^{jk}$ and the definition of the Levi-Civita tensor, we can obtain the Tucker-Hammer equation~\cite{tucker} from the Duffin-Kemmer-Petiau set of equations:
\begin{equation} \left(\begin{array}{cc}
E^2-{\bf p}^2-2m^2&E^2-{\bf p}^2+2E({\bf S\cdot p})+2({\bf S\cdot p})^2\\
E^2-{\bf p}^2-2E({\bf S\cdot p})+2({\bf S\cdot p})^2&E^2-{\bf p}^2-2m^2
\end{array}\right) \left( \begin{array}{c}
\chi\\ \psi
\end{array}\right)=0. 
\label{tucker1}
\end{equation}
In the covariant form the equation (\ref{tucker1}) is written:
\begin{equation}
\left( \gamma^{\mu \nu}p_{\mu}p_{\nu}+p^{\mu}p_{\mu}-2m^2\right)\Psi_{(6)}(p^{\mu})=0. 
\end{equation}
with the $6\times 6$ Barut-Muzinich-Williams matrices~\cite{barut}):
\begin{equation}
\gamma^{00}=\left(\begin{array}{cc}0&1_{3\times 3}\\1_{3\times 3}&0\end{array}\right),\,
\gamma^{i0}=\gamma^{0i}=\left(\begin{array}{cc}0&-S^i\\S^i&0\end{array}\right),\,
\gamma^{ij}=\left(\begin{array}{cc}0&-\delta_{ij}+S_iS_j+S_jS_i\\-\delta_{ij}+S_iS_j+S_jS_i&0\end{array}\right).
\label{gamma}
\end{equation}
In the coordinate space we have
$\left( \gamma^{\mu \nu}\partial_{\mu}\partial_{\nu}+\partial^{\mu}\partial_{\mu}+2m^2\right)\Psi(x^{\mu})=0$.
If we set the condition $\partial_{\mu}\partial_{\mu}\rightarrow -m^2$ we can recover the Weinberg equation, ref.~\cite{weinbergp}:
\begin{equation} \Gamma\pmatrix{\chi\cr\psi}=\left(\begin{array}{cc}
-m^2&m^2+2E({\bf S\cdot p})+2({\bf S\cdot p})^2\\
m^2-2E({\bf S\cdot p})+2({\bf S\cdot p})^2&-m^2
\end{array}\right) \left( \begin{array}{c}
 \chi\\
 \psi
\end{array}\right)=0\,, 
\label{weinberg}
\end{equation}
which is in the covariant form
$(\gamma^{\mu \nu}\partial_{\mu}\partial_{\nu}+ m^2)\Psi(x^{\mu})=0$.
Thus, from what we have seen above, we can conclude that the Duffin-Kemmer-Petiau, Proca, Weinberg and Tucker-Hammer equations are all related one another. 
Let us consider the equation (\ref{tucker1}) as a set of equations for the bivector components in the helicity basis. Then, we have ($p=\mid {\bf p} \mid $):
\begin{equation}
u_{1,\uparrow}=\frac{1}{\sqrt{2}}\left(\begin{array}{c} 
\frac{E+p}{m}\chi_{\uparrow}\\
\frac{m}{E+p}\chi_{\uparrow}\end{array}\right),\,
u_{1,\rightarrow}=\frac{1}{\sqrt{2}}\left(\begin{array}{c} 
\chi_{\rightarrow}\\
\chi_{\rightarrow}\end{array}\right),\,
u_{1,\downarrow}=\frac{1}{\sqrt{2}}\left(\begin{array}{c} 
\frac{m}{E+p}\chi_{\downarrow}\\
\frac{E+p}{m}\chi{\downarrow}\end{array}\right),
\end{equation}
\begin{equation}
v_{1,\uparrow}=\frac{1}{\sqrt{2}}\left(\begin{array}{c} 
\frac{E+p}{m}\chi_{\uparrow}\\
-\frac{m}{E+p}\chi_{\uparrow}\end{array}\right),\,
v_{1,\rightarrow}=\frac{1}{\sqrt{2}}\left(\begin{array}{c} 
\chi_{\rightarrow}\\
-\chi_{\rightarrow}\end{array}\right),\,
v_{1,\downarrow}=\frac{1}{\sqrt{2}}\left(\begin{array}{c} 
\frac{m}{E+p}\chi_{\downarrow}\\
-\frac{E+p}{m}\chi{\downarrow}\end{array}\right),
\end{equation}
where the 3-``spinors" are in the helicity basis  (see~\cite[p.192]{Var}):
\begin{equation}
\chi_{\uparrow}=\left(\begin{array}{c}
\frac{1+\cos\theta}{2}e^{-i\phi}\\
\frac{\sin\theta}{\sqrt{2}}\\
\frac{1-\cos\theta}{2}e^{i\phi}\end{array}\right),\hspace{2mm}
\chi_{\rightarrow}=\left(\begin{array}{c}
-\frac{\sin\theta}{\sqrt{2}}e^{-i\phi}\\
\cos\theta\\
\frac{\sin\theta}{\sqrt{2}}e^{i\phi}\end{array}\right),\hspace{2mm}
\chi_{\downarrow}=\left(\begin{array}{c}
\frac{1-\cos\theta}{2}e^{-i\phi}\\
-\frac{\sin\theta}{\sqrt{2}}\\
\frac{1+\cos\theta}{2}e^{i\phi}\,\end{array}\right).
\end{equation}
The normalization condition is chosen $\chi^{\dagger}\chi=1$.

Now we are ready to study the discrete symmetry operations for the spin-1 case (as we did for the spin-1/2 case in the previous Section). The bivectors have the following properties:
\begin{enumerate} 
\item 
The Parity (${\bf p}\rightarrow -{\bf p}$, $\theta\rightarrow \pi-\theta$, $\phi\rightarrow \pi +\phi$). We note that the 3-``spinors"  are transformed as $\chi_h\rightarrow-\chi_{-h}$; the parity operator is $P=\gamma^{00}$ (it is analogous to that which was used for spin-1/2). Therefore,
\begin{eqnarray}
Pu_{1,\uparrow}(-{\bf p})&=&-u_{1,\downarrow}({\bf p}), \hspace{2mm} 
Pu_{1,\rightarrow}(-{\bf p})=-u_{1,\rightarrow}({\bf p}), \hspace{2mm}
Pu_{1,\downarrow}(-{\bf p})=-u_{1,\uparrow}({\bf p}),\\
Pv_{1,\uparrow}(-{\bf p})&=& +v_{1,\downarrow}({\bf p}), \hspace{2mm} 
Pv_{1,\rightarrow}(-{\bf p})=+v_{1,\rightarrow}({\bf p}), \hspace{2mm}
Pv_{1,\downarrow}(-{\bf p})=+v_{1,\uparrow}({\bf p})\,.
\end{eqnarray}

\item 
The Charge Conjugation is defined $C=e^{i\alpha}\pmatrix{0&\Theta\cr-\Theta&0}{\cal K}$,
with $\Theta_{[1]}=\left(\begin{array}{ccc}0&0&1\\0&-1&0\\1&0&0\end{array}\right)$.
Hence, $\Theta\chi_{\uparrow}^\ast=\chi_{\downarrow}$, $\Theta\chi_{\downarrow}^\ast=\chi_{\uparrow}$, $\Theta\chi_{\rightarrow}^\ast=-\chi_{\rightarrow}$. Finally, we have
\begin{eqnarray}
Cu_{1,\uparrow}({\bf p})&=&+e^{i\alpha}v_{1,\downarrow}({\bf p}),  
Cu_{1,\rightarrow}({\bf p})=-e^{i\alpha}v_{1,\rightarrow}({\bf p}), 
Cu_{1,\downarrow}({\bf p})=+e^{i\alpha}v_{1,\uparrow}({\bf p}),\\
Cv_{1,\uparrow}({\bf p})&=&-e^{i\alpha}u_{1,\downarrow}({\bf p}),  
Cv_{1,\rightarrow}({\bf p})=+e^{i\alpha}u_{1,\rightarrow}({\bf p}), Cv_{1,\downarrow}({\bf p})=-e^{i\alpha}u_{1,\uparrow}({\bf p})\,.
\end{eqnarray}

\item The $CP$ and $PC$ operations:
\begin{eqnarray} CPu_{1,\uparrow}({\bf -p})&=& -e^{i\alpha}v_{1,\uparrow}({\bf p}),\,CPv_{1,\uparrow}({\bf -p})=-e^{i\alpha}u_{1,\uparrow}({\bf p}),\\
CPu_{1,\downarrow}({\bf -p})&=&-e^{i\alpha}v_{1,\downarrow}({\bf p}),\,
CPv_{1,\downarrow}({\bf -p})=-e^{i\alpha}u_{1,\downarrow}({\bf p}),\\
CPu_{1,\rightarrow}({\bf -p})&=& +e^{i\alpha}v_{1,\rightarrow}({\bf p}),\,CPv_{1,\rightarrow}({\bf -p})=+e^{i\alpha}u_{1,\rightarrow}({\bf p}).\end{eqnarray}

\end{enumerate}

We found within the classical field theory that the properties of a particle and an anti-particle of spin-1 are different comparing with the known cases (when the basis is chosen in such a way that the solutions are the eigenstates of the parity).

\section{The Conclusions.}

Similarly to the $({1\over 2},{1\over 2})$ representation~\cite{Grei},
the $({1\over 2},0)\oplus (0,{1\over 2})$ and $(1,0)\oplus (0,1)$ field functions
in the helicity basis are {\it not} the eigenstates of the
common-used parity operator; $\vert {\bf p},\lambda> \Rightarrow
\vert -{\bf p},-\lambda >$  on the classical level. This
is in accordance with the earlier consideration of Berestetski\u{\i},
Lifshitz and Pitaevski\u{\i}. Helicity field functions may satisfy the ordinary Dirac equation
with $\gamma$'s to be in the spinorial representation. 
Helicity field functions can be expanded in the set of the Dirac
4-spinors by means of the matrix ${\cal U}^{-1}$ given in this paper.
$P$ and $C$ operations anticommute  in this framework on the
classical level.
The different formulations of the spin-1 particles are all connected by algebraic transformations.
The properties of spin-1 solutions in the helicity basis with respect to $P$, $C$, $CP$ are similar to those in the spin-1/2 case, and differ from the usual case.

In order to make the above conclusions to be more rigorous one should repeat the calculations in the Fock space within the ``secondary quantization" framework 
(see~\cite{valerihelic} for the spin-1/2 case).

\end{document}